\documentclass[preprint,amsmath,amssymb]{elsarticle}

\begin{document}

\title{Dynamics of genuine multipartite entanglement under local non-Markovian dephasing}
\author{Mazhar Ali}
\address{Department of Electrical Engineering, COMSATS Institute of Information Technology, 22060 Abbottabad, Pakistan}

\begin{abstract}
We study dynamics of genuine entanglement for quantum states of three and four qubits under non-Markovian dephasing. 
Using a computable entanglement monotone for multipartite systems, we find that GHZ state is quite resilient state whereas the 
W state is the most fragile. We compare dynamics of chosen quantum states with dynamics of random pure states and weighted graph states.
\end{abstract}

\maketitle

\section{Introduction}
\label{Sec:intro}

Quantum entanglement not only defies our classical intuition but also finds its role in several practical applications devised to harness 
the power of quantum physics. The technological promise of entanglement has attracted lot of interest to develop a theory of its own, which 
deals with its characterization and quantification, optimal detection in theory and experiments, and the methods to reverse the inevitable 
process of decoherence \cite{Horodecki-RMP-2009, gtreview}. As entangled states are desired to generate and manipulate in experiments, 
therefore it is essential to study the effects of various environments on entanglement. 
In recent years, this study has received considerable attention and is currently an active area of research \cite{Aolita-review}.
The dynamics of entanglement under various environments were studied for both bipartite and multipartite systems
\cite{Yu-work,lifetime,Aolita-PRL100-2008,bipartitedec,Band-PRA72-2005,lowerbounds,Lastra-PRA75-2007, Guehne-PRA78-2008,Lopez-PRL101-2008, 
Ali-work, Weinstein-PRA85-2012}.
Several works of entanglement dynamics considered bipartite aspects of entanglement of multipartite states \cite{Band-PRA72-2005}, however this 
can only give partial results because entanglement in multipartite systems is different than the entanglement among different partitions.
As theory of multipartite entanglement is still in progress, one can make statements on lower bounds of entanglement but not on its exact 
value \cite{lowerbounds}. The exact value of multipartite entanglement was only calculated for specific model of decoherence and specific 
quantum states \cite{Guehne-PRA78-2008}. In addition, in order to comment on the robustness or fragility of a state, one need to compare 
its dynamics with dynamics of random states. Recently, we have addressed these issues and studied the robustness of several multipartite 
states by investigating an exact measure of genuine multipartite entanglement under Markovian environments \cite{Ali-JPB-2014}.

In this work, we extend our study to dynamics of genuine entanglement under a specific type of non-Markovian noise. Although, 
the approximation of weak interaction and no back action of environment on principal system might be valid for certain circumstances, however 
in reality most systems are non-Markovian. It is important to simulate the effects of non-Markovian environments on genuine entanglement. 
Recent progress in the theory of multipartite entanglement has enabled us to study decoherence effects on actual multipartite 
entanglement and not on entanglement among bipartitions. In particular, the ability to compute genuine negativity for multipartite systems 
has eased this task \cite{Bastian-PRL106-2011}. We find that under non-Markovian dephasing, GHZ state appears to be resilient as it repeats 
the collapse and revival of genuine entanglement for long time. On the other hand, the W state turns out to be most fragile state. 
All other quantum states of four qubits exhibit behavior between these two extremes. We compare 
the dynamics of chosen quantum states of three and four qubits with dynamics of random pure states and weighted 
graph states. 

This paper is organized as follows. In section \ref{Sec:Model}, we briefly discuss our model of interest. We review the concept of entanglement 
for multipartite systems in section \ref{Sec:GME} where we also describe the method to compute genuine negativity and discuss the quantum states 
which we study in this work. In section \ref{Sec:res}, we provide results and finally we conclude our work in 
section \ref{Sec:conc}.

\section{Local non-Markovian dephasing model} 
\label{Sec:Model}

The model which we intend to study here is well known \cite{Daffer-PRA70-2004}. Recently, this model has also been studied for sudden 
change in dynamics of quantum discord \cite{Pinto-PRA88-2013}. We consider a dephasing with colored noise with dynamics 
described by a master equation \cite{Daffer-PRA70-2004,Pinto-PRA88-2013} 
\begin{equation}
\dot{\rho}(t) = \mathcal{K} \, \mathcal{L} \, \rho \,, 
\label{Eq:MENM}
\end{equation}
where $\mathcal{K}$ is time-dependent integral operator whose action on a function is defined as 
\begin{equation}
\mathcal{K} \phi = \int_0^t \, dt' \, k(t-t') \phi(t') \, ,  
\end{equation}
with $k(t-t')$ is a kernel which determines the type of environment memory. $\rho$ is density matrix of the principal system and 
$\mathcal{L}$ is the Lindblad super-operator which describes dynamics of principal system as a result of interaction with environment. 
Note that in the absence of $\mathcal{K}$ in Eq.(\ref{Eq:MENM}) one usually get the master equation with Markovian approximation. 
For a concrete example of a system, we may consider the time-dependent Hamiltonian \cite{Daffer-PRA70-2004, Pinto-PRA88-2013}
\begin{equation}
H(t) = \hbar \, \sum_{i = 1}^3 \Gamma_i(t) \, \sigma_i \,,
\end{equation}
where $\sigma_i$ are the Pauli matrices and $\Gamma_i(t)$ are the independent random variables which obey statistics of a random telegraph signal 
defined as $\Gamma_i(t) = a_i \, (-1)^{n_i(t)}$. The random variable $n_i(t)$ has a Poisson distribution with a mean $t/(2 \, \tau_i)$ 
and $a_i$ is an independent random variable taking values $\pm a_i$. This model applies to any two-level quantum system interacting with an 
environment having random telegraph signal noise. As an example, this could describe a two-level atom subjected to a fluctuating laser field 
that has jump type random phase noise \cite{Daffer-PRA70-2004}.

Using von Neumann equation of motion $\dot{\rho} = - ({\rm i}/\hbar) [H, \, \rho]$, we can write the solution for density matrix of 
two level system as 
\begin{equation}
\rho (t) = \rho(0) - {\rm i} \int_0^t \, dk \sum_i \Gamma_i (k) [\sigma_i, \, \rho(k)]\, . 
\end{equation}
Substituting this equation back into von Neumann equation and performing stochastic average, we obtain \cite{Daffer-PRA70-2004}
\begin{equation}
\dot{\rho}(t) = - \int_0^t dt' \sum_k e^{- (\frac{t-t'}{\tau_k}) } a_k^2 \, \big[ \sigma_k , \, [\sigma_k , \, \rho(t')]\big] \,, 
\label{Eq:MENM2}
\end{equation}
where the correlation functions of random telegraph signal 
\begin{equation}
\langle \Gamma_j (t) \Gamma_k(t')\rangle = a_k^2 \, 
e^{-\frac{|t-t'|}{\tau_k}} \, \delta_{jk} 
\end{equation}
have been inserted. It turns out that the dynamical evolution generated by Eq.(\ref{Eq:MENM2}) 
is completely positive when two of the $a_k$ are zero. This would correspond to a physical situation where noise only acts in one 
direction. In particular if $a_1 = a_2 = 0$, and $a_3 = a$, then the dynamics of the system is that of a dephasing with colored noise. The 
Kraus operators describing the dynamics of two-level system are given as \cite{Daffer-PRA70-2004, Pinto-PRA88-2013}
\begin{eqnarray}
 K_1 &=& \sqrt{[1 + \Lambda(\nu)]/2} \, I_2 \nonumber \\
 K_2 &=& \sqrt{[1 - \Lambda(\nu)]/2} \, \sigma_3 \, ,
\end{eqnarray}
where $I_2$ is $2 \times 2$ identity matrix and $\Lambda(\nu) = e^{-\nu} \, \big[ \cos(\mu \nu) + \sin(\mu \nu)/\mu \big] $, with 
$\mu = \sqrt{(4 a \tau)^2 - 1}$ and $\nu = t/(2\tau)$ is the dimensionless time. The Kraus operators satisfy the 
normalization condition $\sum_i \, K_i^\dagger(t) \, K_i(t) =  I$. 

As we are interested in three and four qubit systems, the time evolution of an initial density matrix can be written as
\begin{equation}
\rho(t) = \sum_i  \, M_i(t) \, \rho(0) \, M_i^\dagger(t), 
\label{Eq:TE}
\end{equation}
where $M_i(t)$ are the Kraus operators, satisfying the 
normalization condition $\sum_i \, M_i^\dagger(t) \, M_i(t) =  I$. 
For three qubits, there are $8$ such operators, that is, $M_1 = K_1^A K_1^B K_1^C$, $M_2 = K_1^A K_1^B K_2^C$,$\ldots$, 
$M_8 = K_2^A K_2^B K_2^C$. We have omitted the tensor product symbol between these operators. Similarly, the respective $16$ operators for 
four qubits are $M_1 = K_1^A K_1^B K_1^C K_1^D$, $M_2 = K_1^A K_1^B K_1^C K_2^D$,$\ldots$,$M_{16} = K_2^A K_2^B K_2^C K_2^D$. 

The time evolved density matrix for a single qubit can directly 
be computed and it is given as   
\begin{eqnarray}
\rho (t) = \left( \begin{array}{cc}
\rho_{11} &  \gamma \rho_{12} \\ 
\gamma \rho_{21} & \rho_{22}  
\end{array} \right), 
\end{eqnarray}
where 
\begin{equation}
\gamma = \mu^{-1}\, e^{- \nu } [ \, \sin(\mu \nu) + \mu \, \cos(\mu \nu)\, ].  
\label{Eq:gama}
\end{equation}
For more qubits, the calculation of density matrices is straightforward.

\section{Multipartite entanglement and quantum states}
\label{Sec:GME}

In this section, we briefly review the concept of entanglement in multipartite systems and discuss the particular quantum states which we 
study in this article. We want to emphasize at this stage that material in this section is already known in the literature and we 
cite them appropriately.

\subsection{Genuine multipartite entanglement and multipartite negativity} 

We review genuine multipartite entanglement by considering three 
particles $A$, $B$, and $C$, as the generalization to more parties is straightforward. A state is called separable with respect 
to some bipartition, say, $A|BC$, if it is a mixture of product states with respect 
to this partition, that is, 
$\rho_{A|BC}^{sep} = \sum_j \, q_j \, |\phi_A^j \rangle\langle \phi_A^j| \otimes |\phi_{BC}^j \rangle\langle \phi_{BC}^j|$, 
where $q_j \geq 0$ and $\sum_j q_j = 1$. Similarly, separable states for the two other bipartitions are $\rho_{B|AC}^{sep}$ and 
$\rho_{C|AB}^{sep}$. A state is called biseparable if it can be written as a mixture of states which are separable with respect 
to different bipartitions, that is 
\begin{eqnarray}
 \rho^{bs} = p_1 \, \rho_{A|BC}^{sep} + p_2 \, \rho_{B|AC}^{sep} + p_3 \, \rho_{C|AB}^{sep}\,.
\end{eqnarray}
A state is genuinely multipartite entangled if it is not biseparable. In this paper, we study dynamics of this genuine
multipartite entanglement. 

Recently, it has been worked out to detect and characterize multipartite entanglement by using 
positive partial transpose mixtures (PPT mixtures) \cite{Bastian-PRL106-2011}. 
We recall that a two-party state $\rho = \sum_{ijkl} \, \rho_{ij,kl} \, |i\rangle\langle j| \otimes |k\rangle\langle l|$ is PPT if its 
partially transposed matrix 
$\rho^{T_A} = \sum_{ijkl} \, \rho_{ji,kl} \, |i\rangle\langle j| \otimes |k\rangle\langle l|$ has no negative eigenvalues. It is known 
that separable states are always PPT \cite{peresPPT}. The set of separable states with respect to some partition is therefore 
contained in a larger set of states which has a positive partial transpose for that bipartition. 

The states which are PPT with respect to fixed bipartition may be called $\rho_{A|BC}^{PPT}$, $\rho_{B|AC}^{PPT}$, 
and $\rho_{C|AB}^{PPT}$. We ask whether a state can be written as
\begin{eqnarray}
\rho^{PPTmix} = p_1 \, \rho_{A|BC}^{PPT} + p_2 \, \rho_{B|AC}^{PPT} + p_3 \, \rho_{C|AB}^{PPT}\,.
\end{eqnarray}
Such a mixing of PPT states is called a PPT mixture. 
The genuine multipartite entanglement of four or more particles 
can be detected and quantified in an analogous manner by considering 
all bipartitions (like one party vs. $N-1$ parties, 
two parties vs. $N-2$ parties, etc.).

As any biseparable state is a PPT mixture, therefore any state which is not a PPT mixture is guaranteed to be genuinely 
multipartite entangled. The prime advantage of considering PPT mixtures instead of biseparable states is that 
PPT mixtures can be fully characterized with the method of semidefinite programming (SDP) \cite{sdp}. In general, the set of 
PPT mixtures is a very good approximation to the set of biseparable states and delivers the best known separability criteria for 
many cases, nevertheless there are multipartite entangled states which are PPT mixtures \cite{Bastian-PRL106-2011}. It is also 
interesting to note that there are biseparable states which may have negative partial transpose (NPT) under each partition \cite{gtreview}.

It has been shown \cite{Bastian-PRL106-2011} that a state is a PPT mixture iff the following optimization problem 
\begin{eqnarray}
 \min {\rm Tr} (\mathcal{W} \rho)
\end{eqnarray}
under the constraint that for all bipartitions $M|\bar{M}$
\begin{eqnarray}
 \mathcal{W} = P_M + Q_M^{T_M},
 \quad \mbox{ with }
 0 \leq P_M\,\leq I \mbox{ and }
 0 \leq  Q_M  \leq I\, 
\end{eqnarray}
has a positive solution. The constraints reflect that $\mathcal{W}$ is a decomposable entanglement
witness for any bipartition. If this minimum is negative then $\rho$ is not a PPT mixture and hence genuinely multipartite 
entangled. Since this is a semidefinite program, the minimum can be efficiently computed and the optimality of the solution can
be certified \cite{sdp}. We use the programs YALMIP and SDPT3 \cite{yalmip} to solve SDP. We also use implementation which is freely 
available \cite{PPTmix}. 

This approach can be used to quantify genuine entanglement as the absolute value of the minimization was shown 
to be an entanglement monotone for genuine multipartite entanglement \cite{Bastian-PRL106-2011}. In the following, 
we will denote this measure by $E(\rho)$. For bipartite systems, this monotone is equivalent to {\it negativity} \cite{Vidal-PRA65-2002}. 
For a system of qubits, this measure is bounded by  $E(\rho) \leq 1/2$ \cite{bastiangraph}.

\subsection{Multipartite entangled states}
\label{Sec:RIQS}

We are interested in several families of states in this work. Two important families of states, namely the GHZ states and the W states 
for $N$ qubits are given as  
\begin{eqnarray}
|GHZ_N \rangle &=& \frac{1}{\sqrt{2}}(|00...0\rangle + |11...1\rangle), \nonumber \\
|W_N\rangle &=& \frac{1}{\sqrt{N}}(|00...001\rangle + |00...010\rangle + \ldots + |10...000\rangle).
\label{Eq:GHZ3Qb1}
\end{eqnarray}
GHZ state has always maximum value of monotone, that is, $E(|GHZ_N\rangle\langle GHZ_N|) = 1/2$, whereas for the W state, numerical 
value depends on the number of qubits. For three qubits $E(|W_3\rangle\langle W_3|) \approx 0.443$ and for four qubits 
$E(|W_4\rangle\langle W_4|) \approx 0.366$.

Several interesting states for four qubits are Dicke state $|D_{2,4}\rangle$, the singlet state $|\Psi_{S,4}\rangle$, the cluster 
state $|CL\rangle$ and the so-called $\chi$-state $|\chi_4\rangle$, given as 
\begin{eqnarray}
|D_{2,4} \rangle &=& \frac{1}{\sqrt{6}} [ |0011 \rangle + |1100\rangle + |0101 \rangle + |0110\rangle + |1001 \rangle 
+ | 1010\rangle] \,,  \nonumber \\
|\Psi_{S,4}\rangle &=& \frac{1}{\sqrt{3}} [ |0011 \rangle + |1100\rangle - \frac{1}{2} ( \, |0101 \rangle + |0110\rangle  
+ |1001 \rangle + |1010\rangle)] \,, \nonumber \\
|CL \rangle &=& \frac{1}{2} [|0000 \rangle + |0011\rangle + |1100 \rangle - |1111\rangle], \nonumber \\
|\chi_4 \rangle &=& \frac{1}{\sqrt{6}} [\sqrt{2} | 1111 \rangle + |0001 \rangle + |0010\rangle + |0100 \rangle + |1000\rangle], 
\end{eqnarray}
respectively. All these states are maximally entangled with respect to multipartite negativity, 
$E (|D_{2,4}\rangle \langle D_{2,4}|) = E (|\Psi_{S,4}\rangle \langle \Psi_{S,4}|) = E (|CL\rangle \langle CL|) 
= E (|\chi_{4}\rangle\langle \chi_4|) = 1/2$. These states along with their properties are discussed in Ref.~\cite{gtreview}.

\subsection{Random pure states}

We describe the generation of random pure states. A state vector randomly distributed according to the Haar measure  
can be generated as follows \cite{Toth-Arxiv}: First, we generate a vector such that both the real and the 
imaginary parts of the vector elements are Gaussian distributed random numbers with a zero mean and unit 
variance. Second we normalize the vector. It is easy to prove that the random vectors obtained this way are equally distributed 
on the unit sphere \cite{Toth-Arxiv}. We stress that we generate random pure states in the 
global Hilbert space of three- and four qubits, so the unit sphere is not the Bloch ball.

\subsection{Weighted graph states}

Another important family of multi-qubit states are weighted graph states which also includes states 
such as GHZ and cluster states \cite{Hartmann-JPB40-2007, Hein-arXiv2006}. These states have been studied for 
entanglement properties of spin gases \cite{Hartmann-JPB40-2007}. 

We consider a graph as a set of vertices and edges, where the vertices may denote the physical systems (qubits) and the edges 
represent the interactions among physical systems. We initially prepare all the qubits in the state 
$|+\rangle = (|0\rangle + | 1\rangle )/\sqrt{2}$. For any pair of qubits $k,l$ which are connected with an edge, we apply an interaction 
via Hamiltonian
\begin{eqnarray}
H_{kl}  = \frac{1}{4} \, (I - \sigma_z^{(k)}) \otimes (I - \sigma_z^{(l)}) \, .
\end{eqnarray}
The resulting unitary transformation is $U_{kl} = {\rm e}^{−i \, \phi_{kl} \, H_{kl}}$, where $\phi_{kl}$ is the interaction 
time. The generated state after this process is called the weighted graph state given as
\begin{eqnarray}
|WGS\rangle = \bigotimes_{k,l} U_{kl} (\phi_{kl}) \, |+\rangle^{\otimes N} \, .
\end{eqnarray}
The weighted graph state is uniquely determined by the $N(N-1)/2$ parameters $\phi_{kl}$, which form only a small subset of all 
pure states described by $2^N-1$ parameters. However, many interesting belong to this class. We have chosen the interaction 
times $\phi_{kl} \in [0, 2 \pi]$ uniformly distributed in the interval to generate these states. For the choice 
$\phi_{kl} = \pi$ or $\phi_{kl} = 0$, the usual graph states (also containing the GHZ and cluster states) appear. We note that 
the temporal order of the interaction does not matter because the unitaries $U_{kl}$ commute. Some generalizations of these  
states have been investigated recently \cite{LME}.

We point out here that all random states generated by these two techniques are genuinely entangled. As described above, the weighted 
graph states may include GHZ states and cluster states, however this does not mean that every random state generated is equivalent to 
only these two types of states. Therefore, it is meaningful to compare the dynamics of specific states with dynamics of 
random states \cite{Ali-JPB-2014}.  

\section{Results}
\label{Sec:res}

In this section, we present our results for quantum states discussed in previous section. First we discuss the effect of 
non-Markovian dephasing on genuine entanglement of three qubits. 

\begin{figure}[t!]
\scalebox{2.0}{\includegraphics[width=1.95in]{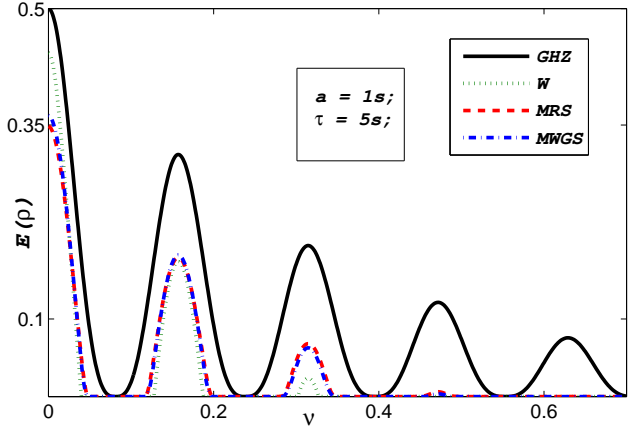}}
\centering
\caption{Multipartite negativity for various three-qubit states is plotted against parameter $\nu$. We observe different rates of 
collapse and revival of genuine entanglement for different states. GHZ state is resilient state whereas W state is fragile.}
\label{Fig:E3QbNMD1}
\end{figure}

Figure \ref{Fig:E3QbNMD1} shows the multipartite negativity $E(\rho)$ plotted 
against the dimensionless time $\nu = t/(2 \tau)$. We have chosen $a = 1s$ and $\tau = 5s$ which are for non-Markovian 
region \cite{Pinto-PRA88-2013}. 
The solid line denotes the GHZ state whereas dotted line is for the W state. It can be seen that the genuine entanglement of GHZ state 
first goes to zero and then revives again with a lower peak than the previous one. This collapse and revival of genuine entanglement for 
GHZ state repeats itself for sufficiently long time, however each time with a lower peak than the earlier ones. This suggest that GHZ state 
is resilient state against non-Markovian dephasing. This observation is similar to our recent studies where we have observed that GHZ state 
is quite robust under Markovian dephasing \cite{Ali-JPB-2014}. We stress that in our previous study \cite{Ali-JPB-2014}, 
we have used logarithmic derivative of genuine entanglement to claim the robustness of GHZ state under Markovian dephasing, whereas 
here we use the term "resilience" to this several time repetition of collapse and revival of genuine entanglement for GHZ state. 
Several authors have studied the robustness of entanglement \cite{Borras-PRA79-2009,Novotny-PRL107-2011}. In \cite{Borras-PRA79-2009}, 
the authors have identified most robust states under local decoherence used the definition introduced before \cite{Band-PRA72-2005}, 
whereas, the authors of Ref.\cite{Novotny-PRL107-2011} considered asymptotic long-time dynamics of initial states and identified two 
different classes of states, one class is fragile even there remains some coherence in the system and the second class as most robust 
states which become disentangled only when decoherence is perfect. However, as described above, our approach is different than these 
studies.

In contrast, the W state is very fragile under non-Markovian dephasing. The collapse and revival of genuine entanglement 
for W state disappear earlier. The revival of genuine entanglement for the W state also takes a little longer time than the GHZ state. 
This observation is similar to Markovian dephasing where the W state is found to be quite fragile \cite{Ali-JPB-2014}. 
In addition, we plot the mean values of genuine negativity for random pure states (MRS) denoted by dashed line and weighted graph 
states (MWGS) denoted by dashed-dotted line for comparison purpose. It is interesting to note that genuine entanglement behavior of 
both type of random states almost resemble to the W state. At this stage, we want to remind the readers that our method to 
detect the genuine entanglement is not always perfect. If this measure is positive then the state is guaranteed to be
genuinely entangled, however, if it is zero, then in general, we do not know whether the state is entangled or not. Except for 
the GHZ state which we discuss below in detail, we can not say with certainty that all zero values of genuine negativity in 
Figure \ref{Fig:E3QbNMD1} and all subsequent figures correspond to states with no entanglement of any type. 

\begin{figure}[t!]
\scalebox{2.0}{\includegraphics[width=1.95in]{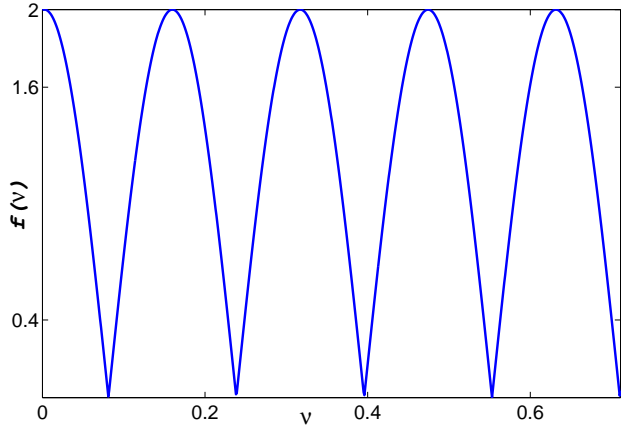}}
\centering
\caption{Absolute value of function $f(\nu)$ scaled down to a factor of $10$, is plotted against parameter $\nu$.}
\label{Fig:IE}
\end{figure}

An interesting property of the dynamical process is the fact that all zero elements of the initial density matrix remain zero. For GHZ state 
the only non-zero density matrix elements are $\rho_{11}$, $\rho_{18}(t)$, $\rho_{81}(t)$ and $\rho_{88}$. A recent result on the detection 
of genuine entanglement states that for biseparable states, the inequality 
\begin{equation}
 |\rho_{18}| \leq \sqrt{\rho_{22}\rho_{77}} + \sqrt{\rho_{33}\rho_{66}} + \sqrt{\rho_{44}\rho_{55}}
\end{equation}
is satisfied and the violation implies genuine entanglement \cite{Otfried-NJP12-2010}. This criterion is a necessary and 
sufficient condition for GHZ-diagonal states \cite{Otfried-NJP12-2010}. We will not go into details of GHZ-diagonal states here but for our 
purpose this criterion would imply that time evolved GHZ state is genuinely entangled if and only if $|\rho_{18}(t)| > 0$, that is,
\begin{equation}
|\gamma^3 /2 | > 0, \quad \Longrightarrow \, f(\nu) = |\sin(\mu \nu) + \mu \cos(\mu \nu)| > 0 \,,  
\end{equation}
where $\gamma$ is defined in Eq.(\ref{Eq:gama}). In Figure \ref{Fig:IE}, we plot the function $f(\nu)$ scaled down to a factor of $10$ 
against parameter $\nu$ with $\mu = \sqrt{399}$ for chosen values of $a$ and $\tau$. A close comparison of Figure \ref{Fig:IE} with 
Figure \ref{Fig:E3QbNMD1} reveals that for those points where $f(\nu)$ is zero, the genuine entanglement for GHZ state is also zero and vice 
versa. Actually, at these points the off-diagonal elements of any arbitrary time-evolved density matrix disappear completely, so 
entanglement of any type disappear as well. This fact is evident in Figure \ref{Fig:E3QbNMD1}. 
It also explains the collapse and revivals of genuine entanglement of GHZ state at these instances.
It was shown \cite{Filippov-PRA85-2012} that phase-damping channels are called locally entanglement-annihilating if and only if the 
off-diagonal elements disappear completely. This happens only at times when $f(\nu) = 0$. Therefore, only at these points, being 
entanglement-annihilating coincides with being entanglement-breaking. It is interesting to note that disappearance of 
genuine entanglement only coincides with these instances for GHZ state. For other states, there are intervals where off-diagonal 
elements are not zero but our criterion fails to detect whether there is entanglement of any type as discussed before.

\begin{figure}[t!]
\scalebox{2.0}{\includegraphics[width=1.95in]{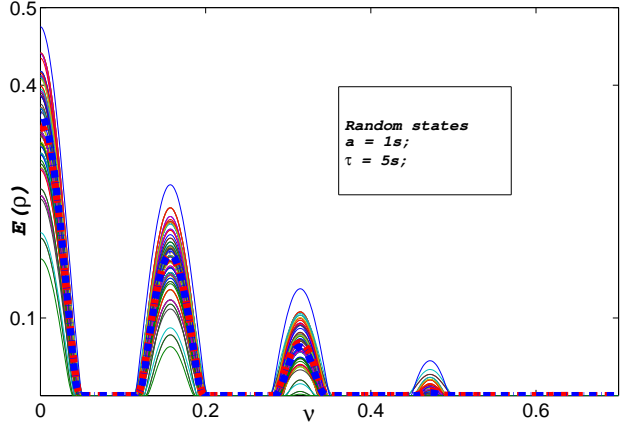}}
\centering
\caption{Genuine entanglement for three-qubit random states with mean values of random states and weighted graph states 
plotted against parameter $\nu$.}
\label{Fig:E3QbRSNMD}
\end{figure}

\begin{figure}[t!]
\scalebox{2.0}{\includegraphics[width=1.95in]{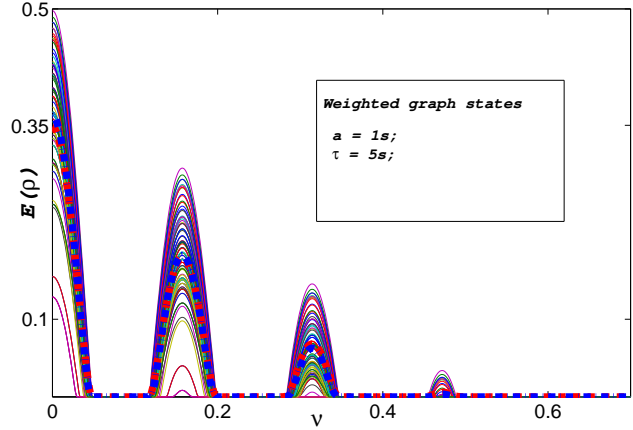}}
\centering
\caption{Genuine entanglement for three-qubit weighted graph states with mean values of random states and weighted graph states 
plotted against parameter $\nu$.}
\label{Fig:E3QbWGSNMD}
\end{figure}

The presence of exponential factor in the definition of $\gamma$ also explains the gradually decaying peaks of genuine entanglement of 
all initial states. As every off-diagonal element of the time-evolved density matrix is being multiplied by a factor $\gamma^r$ with $r = 1$, 
$2$, or $3$, therefore as $f(\nu)$ is zero necessarily means $\gamma$ is zero and $\gamma$ is maximum when $f(\nu)$ is maximum. 
This is the reason that all states have their peak value of genuine entanglement at instances when $f(\nu)$ is maximum. 
This feature is also evident from Figures \ref{Fig:E3QbRSNMD} and \ref{Fig:E3QbWGSNMD}.

\begin{figure}[t!]
\scalebox{2.20}{\includegraphics[width=1.95in]{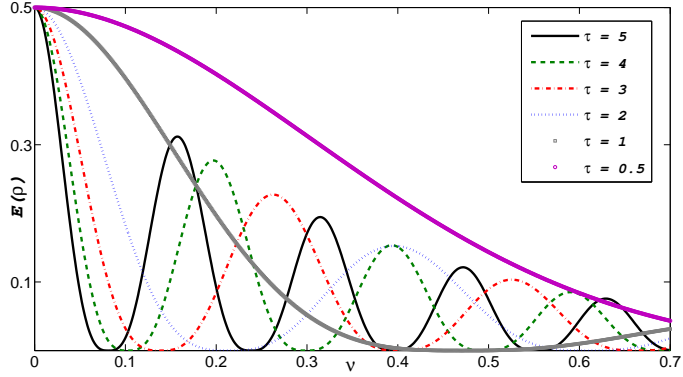}}
\centering
\caption{Genuine entanglement for three-qubit GHZ state is plotted against parameter $\nu$. We observe that by decreasing degree of 
non-Markovianity, collapse and revival of genuine entanglement are also decreased and for $\tau = 0.5$, which corresponds to Markovian case, 
there are no collapse and revivals.}
\label{Fig:E3ghzNMD}
\end{figure}
As the behavior of GHZ state is very different than all other states, therefore in order to get some more insight, we explore the effects 
of decreasing the degree of non-Markovianity on its dynamics. By fixing $a = 1s$, the parameter $\tau = 5$ denotes the non-Markovian case, 
whereas $\tau = 0.5$ corresponds to Markovian case. Figure \ref{Fig:E3ghzNMD} shows genuine entanglement of GHZ state for various values of 
parameter $\tau$ plotted against parameter $\nu$. It can be observed that as we decrease the degree of non-Markovianity, the collapses and 
revivals of genuine entanglement are also decreased and delayed. Finally, for $\tau = 0.5$, there are no collapse and revival of genuine 
entanglement, as we expect from Markovian dynamics. As we discussed above, for GHZ state to be genuinely entangled, $|\rho_{18}(t)| > 0$, and 
this value could only be non-zero if $\gamma$ defined in Eq.(\ref{Eq:gama}) is non-zero. As $\gamma$ depends on $\mu$, which in turn depends 
on $\tau$, that is, for $\tau = 5$, $\mu = \sqrt{399}$ (non-Markovian case), for such a large value of $\mu$, the sine and cosine functions 
are considerable and can not be ignored. Hence these functions are responsible for collapses and revivals. On the other hand, for $\tau = 0.5$, 
we have $\mu = \sqrt{3}$ (Markovian case), which is much smaller than the previous case. So for such a small argument, we can apply small angle approximation 
to sine and cosine functions. Hence we can explain the corresponding disappearance of collapses and revivals of genuine entanglement 
for Markovian case.

Figure \ref{Fig:E3QbRSNMD} shows multipartite negativity for $100$ random pure states plotted against $\nu$. We also plot mean values of 
random pure states (MRS) denoted by dashed line and weighted graph states (MWGS) denoted by dashed-dotted line, however they almost 
overlap and may not be clearly visible. 

In Figure \ref{Fig:E3QbWGSNMD}, we plot multipartite negativity for $100$ weighted graph states along with mean values of genuine 
entanglement for random pure states (MRS) and weighted graph states (MWGS).

\begin{figure}[t!]
\scalebox{2.0}{\includegraphics[width=1.95in]{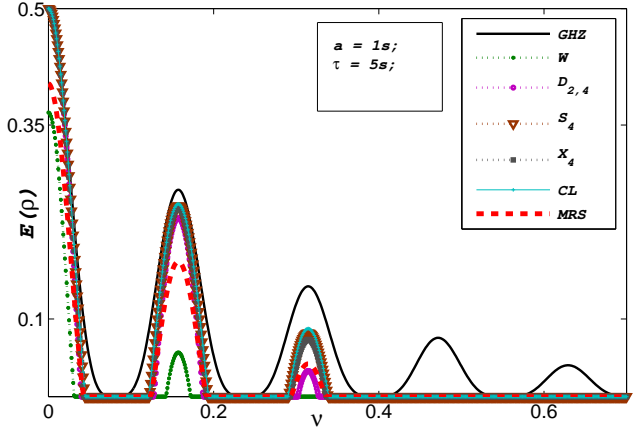}}
\centering
\caption{Genuine entanglement for various four-qubit states with mean value of random pure states is plotted against parameter $\nu$.}
\label{Fig:E4QbNMD}
\end{figure}

Let us now discuss the results for four qubits case. In Figure \ref{Fig:E4QbNMD}, we plot genuine negativity for various states discussed in 
previous section. As for the 
case of three qubits the GHZ state is resilient state whereas the W state is quite fragile. All other states including random pure 
states and weighted graph states exhibit a trend in between these two extremes as evident from the Figure \ref{Fig:E4QbNMD}. 
As results for the four qubits are almost identical to the three qubits case, therefore we preferred to 
plot only mean value of genuine entanglement for random states only.

\section{Conclusion}
\label{Sec:conc}

We have studied the behavior of genuine multipartite entanglement under non-Markovian dephasing. 
Using a computable entanglement monotone for multipartite quantum systems, we have observed the collapses and revivals of 
genuine entanglement for various quantum states of three and four qubits. We have found that GHZ state is resilient state as it repeats its 
revival for long time whereas all other states loose their revivals much earlier. We have examined and explained this different behavior of 
GHZ state. We have found that the W state is the most fragile state similar  
to Markovian environments. We have compared dynamics of chosen quantum states with dynamics of random pure states and weighted graph states 
so that we can make meaningful statements about their behavior under decoherence. We found that all random states and weighted graph states 
show a similar trend as the W state. We stress here that our conclusions are based on a criterion whose positive value for a given 
quantum state is guaranteed to be genuinely entangled, however, for states which are not detected by this criterion, we are not certain about 
their entanglement properties. For GHZ state under current dynamics, this criterion provides a necessary and sufficient criterion to 
detect genuine entanglement.   

\section*{Acknowledgements}

The author is grateful to both anonymous referees for their constructive comments, which brought much clarity and a new 
Figure \ref{Fig:E3ghzNMD} in the manuscript.

\section*{References}

\end{document}